# A Mechanistic Study on Environment Gases in Metal Additive Manufacturing

Zhongshu Ren[1,2,3,*], Samuel J. Clark[4], Lin Gao[2,4], Kamel Fezzaa[4], Tao Sun[1,2,*]

[1] Department of Mechanical Engineering, Northwestern University, Evanston, IL 60208.

[2] Department of Materials Science and Engineering, University of Virginia, Charlottesville, VA 22904.

[3] NSLS-II, Brookhaven National Laboratory, Uptown, NY 11973

[4] Argonne National Laboratory, Lemont, IL 60439.

* Corresponding authors' emails: taosun@northwestern.edu, zren2@bnl.gov

**Abstract**: A variety of protective or reactive environmental gases have recently gained growing attention in laser-based metal additive manufacturing (AM) technologies due to their unique thermophysical properties and the potential improvements they can bring to the build processes. However, much remains unclear regarding the effects of different gas environments on critical phenomena in laser AM, such as rapid cooling, energy coupling, and defect generation. Through simultaneous high-speed synchrotron x-ray imaging and thermal imaging, we identify distinct effects of various environmental gases in laser AM and gained a deeper understanding of the underlying mechanisms. Compared to the commonly used protective gas, argon, it is found that helium has a negligible effect on cooling the part. However, helium can suppress unstable keyholes by decreasing effective energy absorption, thus mitigating keyhole porosity generation and reducing pore size under certain processing conditions. These observations provide guidelines for the strategic use of environmental gases in laser AM to produce parts with improved quality.

**Keywords:** laser powder bed fusion, multimodal characterization, gas environment

# 1. Introduction

Metal additive manufacturing (AM), such as laser powder bed fusion (LPBF), has seen rapid growth in applications for producing end-use parts [1-3]. The disruption of the global supply chain during the pandemic has accelerated the resurgence of this technology, which is revolutionizing the manufacturing industry. Metal AM offers numerous advantages over conventional manufacturing techniques, including the ability to print complex parts, shorten supply chain, enable high customization, facilitate on-site and on-demand production, and reduce energy consumption [4].

In a typical LPBF process, a high-power laser melts a thin layer of metal powder in selective locations according to the computer design. This melting and solidification process repeats layer by layer to produce a 3D part. Due to the high temperature and fast-heating and cooling cycle, the LPBF process exhibits intricate phenomena [5, 6]. As the metal absorbs the laser energy, the high temperature can evaporate the metal and generates a strong recoil pressure, pushing down the liquid metal and forming a vapor depression (aka, keyhole) [7-10]. Away from the keyhole region, where the temperature is lower, a melt pool is formed, inside which complex melt flow present. These complex phenomena may lead to poor part quality if not controlled well, causing catastrophic failure of parts which are highly undesired in risk-averse sectors such as aerospace, energy, biomedical, and defense [11, 12].

Inert gases are often used in the printing chamber to prevent chemical reactions (e.g. oxidation) of the processed metal. Argon has long been the primary environment gas, but other alternative gases are gaining increased attention [13, 14]. Helium, for example, is a ambient gas with distinct thermophysical property compared to argon. Despite its higher cost, which limits its widespread use in industry, helium has the potential to improve certain spects of the printing process. Due to differences in properties such as density and thermal conductivity, helium is thought to offer several advantages over argon. For example, compared to argon, helium has been observed to generate less spatter [15], which is believed to contribute to a smoother surface finish [16].

However, much remains unclear regarding the effect of enviroment gases on the printing process and the underlying mechanisms. This has led to some confusion and debate in the AM community:

- The effect of envionment gases on the cooling rate
- The effect of envionment gases on energy coupling
- The effect of envionment gases on voids formation

Amando et al simulated the LPBF process and found that helium has a faster cooling rate compared to argon [17]. In contrast, Pauzon et al didn't observe such a faster cooling effect by helium in LPBF process from their x-ray diffraction results [18]. Regarding the debate revolving around the laser-metal energy coupling, Caballero et al found a distinct trend in the curve of melt track vs energy density [19]. They attributed that to the smaller plasmas generated in the helium environment, alleviating the laser-plasma interaction, implying an influenced the laser absorption by the metal. On the other hand, Traore et al argued that the 0.1-1.0 kW Yb:YAG laser commonly used in LPBF was too weak to influence the energy coupling, owing to its short wavelength and consequent much smaller absorption coefficient [15]. Instead, they found that the total depth of the melt track is comparable between the helium and argon environments and concluded an unchanged energy-coupling in helium. It is important to note that both research teams used the ex situ cross-sectional microscopy to characterize the as-printed samples. Additionally, Pauzon et al reported a decrease in lack-of-fusion voids in the sample printed in a helium environment than that in argon [20]. However, in a following work, the same research team reported a different result measured using x-ray computed tomography: an increased lack-of-fusion voids samples processed in helium [21].

In the present work, we conducted a systematic study to examine the mechanistic effects of environment gases on the LPBF process with more sophisticated in situ characterization. High-speed synchrotron x-ray imaging and simultaneous high-speed near-infrared (NIR) imaging were performed to directly compare the printing process in different gas environments (argon versus helium). Through this study, we aim to settle the ongoing debates regarding the effects of environment gases, clarify some confusions, and enphasize the importance of carefully examining the role of environment gases, as the process involves complex multi-phase interactions.

## 2. Materials and methods

### 2.1 Materials

We used two different types of samples in this work: Ti-6Al-4V plate samples and Al6061 single-layer powder bed samples. The Ti-6Al-4V plate samples were cut from commercial products (Grade 5, McMaster-Carr, USA) using electro-discharge machining (EDM) and then sanded to the size of ~50 mm×3.00 mm×0.40 mm using sandpaper up to 1200-grit. Each Al6061 powder bed sample consist of a plate as the substrate and a thin layer of powder atop, both of which were sandwiched between two pieces of glassy carbon plates (Grade 22, Structure Probe Inc., USA) (Fig. 1a). The plate substrates were cut from Al6061 stock (McMaster-Carr, USA) and then sanded to ~50 mm×3.00 mm×1.00 mm. The Al6061 powder (Valimet, Inc.) had a particle size of 17~52 μm, while the powder bed had a thickness of ~100 μm.

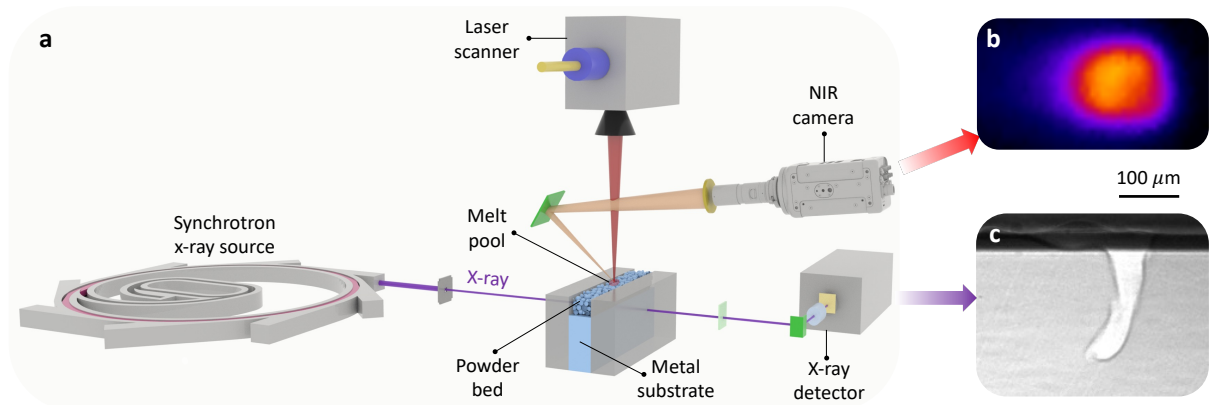

**Fig. 1. Simultaneous *operando* high-speed synchrotron x-ray and NIR imaging of LPBF. a.** Schematic of the experiment setup. High-energy x-ray generated by the undulator penetrated the sample and was captured by the x-ray detector, revealing inner structural dynamics during the printing process. Simultaneously, thermal emissions from the sample surface were recorded by a NIR camera. **b.** A representative NIR image of the sample surface. Thermal information of the melt pool, as well the following solidification process can be quantified from NIR image series. **c.** A representative x-ray image of sample subsurface. The morphologies of the keyhole and melt pool, as well as keyhole porosity, can be visualized. The scale bar applies to both NIR image (panel **b**) and x-ray image (panel **c**).

*2.2 Laser powder bed fusion*

The laser source was a 1070 nm ytterbium fiber laser (IPG YLR-500-AC, USA), which was fed into a galvanometer scanner (intelliSCANde 30, SCANLAB GmbH, Germany) [22]. This Gaussian laser had a spot size of ~77 μm (1/e2) on the sample surface, as characterized by a beam profiler (FBP-2KF Cinogy Technologies GmbH, Germany). During the experiment, after loading the sample into the build chamber, we purged the chamber and then filled it with an environment gas (either argon or helium) back to atmospheric pressure. A single straight line was scanned on the sample by the continuous-mode laser, utilizing a specified power and scan speed.

*2.3 Operando high-speed x-ray imaging*

*Operando* high-speed x-ray imaging was performed at the 32-ID-B beamline of the Advanced Photon Source at Argonne National Laboratory [22, 23]. An undulator with an 18 mm period was set to a gap of 12 mm to produce a white beam with the first harmonic energy at ~24 keV. The x-ray imaging system consisted of a 100-μm thick Lu3Al5O12:Ce scintillator, a 45° reflection mirror, a 10× objective lens (NA = 0.28, Mitutoyo Corp., Japan), a tube relay lens, and an optical high-speed camera. The camera (Photron FastCam SA-Z, Photron Inc., Japan) operated at a frame rate of 50 kHz with an exposure time of 5 μs, achieving a spatial resolution of ~2.0 μm/pixel.

*2.4 Simultaneous high-speed near-infrared (NIR) imaging*

High-speed NIR imaging was conducted simultaneously with the x-ray imaging [24]. The NIR imaging system comprised a wide-band reflection mirror (protected silver mirror, Thorlabs Inc., USA), a 1070 nm notch filter (Edmund Optics Inc., USA), a 760 nm long-pass filter (Newport Corp., USA), and a high-speed camera (FASTCAM NOVA S9, Photron Inc., Japan) equipped with a Resolv4K zoom lens (Navitar, Inc., USA). The camera operated at a frame rate of 200 kHz and an exposure time of 0.3 μs, achieving a spatial resolution of ~9 μm/pixel. The NIR imaging was performed at a view-angle of ~50° relative to the sample surface.

*2.5 Image processing and data analysis*

All x-ray images were processed and analyzed using the image processing software ImageJ [25]. For each image sequence (a series of images in consecutive timestep), an x-ray image associate with the moment before the laser enters the field of view (FOV) was background corrected from the image sequence to enhance the contrast. Keyhole and melt pool morphologies were then extracted from the background corrected image sequence.

Two types of analyses were performed on NIR images using MATLAB R2023b (MathWorks, USA) [24]. The first type of analysis began by extracting a 1D time-series signal from NIR images through thresholding the intensity values to isolate the keyhole area. Then, a continuous wavelet transform (MATLAB function, cwt) was applied to the time-series signal to convert it into a wavelet scalogram, akin to a short-window fast Fourier transform (SWFFT). The scalogram provides valuable insights into the oscillation frequency over localized time intervals. The second type of analysis involved inserting a stationary virtual probe into the NIR image sequence to obtain the time-series cooling curve at the probe's location.

## 3. Results

*3.1 Harmonizing subsurface melting dynamics with surface thermal signatures*

We conducted operando characterizations of the LPBF processes of Ti-6Al-4V plate and Al6061 powder bed samples under two distinct environment gas conditions (helium vs. argon) at 32-ID-B of Advanced Photon Source at argonne National Laboratory. A 1070 nm fiber laser with a Gaussian profile was focused on the sample surface and scanned (in continuous-wave mode) along a single straight line using a galvanometer scanner (Fig. 1a). Absent in the schematic, in the actual experimental setup, there was a chamber enclosing the sample. During each experiment, only one type of environment gas (either helium or argon) filled the chamber, creating a well-controlled comparison experiment condition.

During the laser-scan process, a high-energy x-ray beam (1st harmonic of ~24 keV), generated from the synchrotron x-ray source, passed through the sample and entered the x-ray detector, providing full-field x-ray images (Fig. 1c). With high spatial (2 $\mu$m/pixel) and temporal resolution (5 $\mu$s), as well as a high frame rate (50 kHz frames per second), we quantified keyhole and melt pool morphologies, and keyhole porosities from such images. Meanwhile, a top-view (50° view-angle) near-infrared (NIR) camera was utilized to capture the surface thermal information of the melting process. The spatial and temporal resolutions of the NIR imaging were 9 $\mu$m/pixel and 0.3 $\mu$s, respectively, with a frame rate of 200 kHz, high enough to adequately sample and resolve the characteristic periodical keyhole oscillations (Fig. 1b). This coupled advanced characterization furnished us with comprehensive insights into and quantifiable dataset of the unique combinations of subsurface melting dynamics and surface thermal signatures of the LPBF process. Multimodal characterizations like this have particular importance in gathering comprehensive information, which is ideal for data-driven approaches [10, 26, 27].

*3.2 Keyhole porosity*

A well-defined sharp boundary was identified in the map of laser power (P) and scan speed (V), delineating two distinct keyhole conditions for Ti-6Al-4V samples processed in different gas environments [25]. On the lower right side of the porosity boundary in the P-V map lies the regime of stable keyhole condition, characterized by low energy density, shallow keyhole depth, and most importantly, the absence of keyhole porosity. In contrast, on the upper left side of the porosity boundary in the P-V map, the keyhole condition becomes unstable. Here, the laser energy density is relatively high, leading to deeper keyhole depths and the generation of keyhole porosity. As one moves from the right to the left away from the keyhole boundary, there is an increase in keyhole size accompanying the escalation of keyhole instability (Fig. 1a). Notably, the size of porosity under helium is smaller than that under argon for the same P-V settings. More importantly, the porosity boundary shifts to the left for helium, allocating more P-V space to stable keyhole conditions. The reduced keyhole porosity size and the leftward shift of the porosity boundary in helium environment suggest that helium stabilizes the keyhole, thereby mitigating the generation of keyhole porosity.

Keyhole porosity generation is closely correlated with keyhole oscillations, which have been quantified and visualized through wavelet analysis of the NIR images [24]. The scalogram produced by wavelet analysis illustrates the oscillation intensity across both frequency and time domains. Single-modal intrinsic oscillations, associated with a stable keyhole, do not generate keyhole porosity. These are characterized by bright horizonal narrow bands persisting over time, observable at V = 600 mm/s in argon and 500~600 mm/s in helium (Fig. 2b). The presence of keyhole porosity was examined through simultaneous x-ray imaging (Fig. 3a, and Supplementary Movies 1-6). High frequency components (perturbative oscillations) within the multimodal oscillations are identified as culprits for keyhole porosity generation, due to the frequent formation of protrusions at front and back keyhole walls and consequent keyhole collapse (400~500 mm/s in argon and 400 mm/s in helium in Fig. 2b). Under either gas condition, a transition from high to low V reduces the intrinsic frequency, and crossing the boundary triggers the onset of perturbative oscillations. The influence of helium is evidenced by a direct comparison between argon and helium at a 500 mm/s scan speed (Fig. 2b). This supports our previous findings that helium plays a stabilizing role in keyhole oscillations.

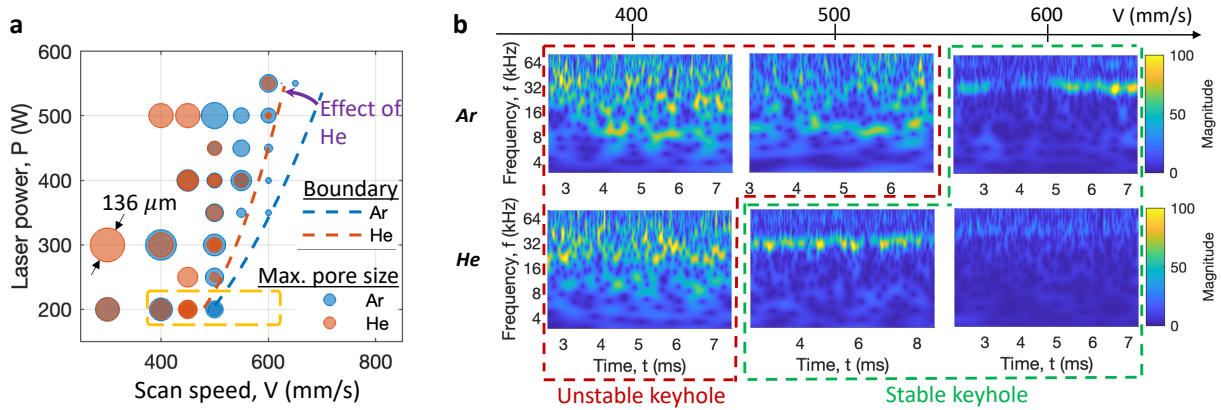

**Fig. 2. Effect of environment gases on keyhole dynamic in laser melting of Ti-6Al-4V plates. a.** P-V map demonstrating the shift in the keyhole porosity boundary (from blue to orange dashed curve) by replacing argon with helium. A keyhole porosity boundary separates the stable keyhole regime (positioned on the right lower side of the boundary, characterized by the absence of keyhole porosity) from the unstable keyhole regime (located on the left higher side of the boundary, where keyhole pores are present). Within the unstable keyhole regime, the diameters of the blue and orange circles represent the measured maximum equivalent pore size for each P-V combination. **b.** Tableau of representative wavelet scalograms from NIR images at a constant laser power (P) of 200 W and varying scan speed (V) across the keyhole porosity boundary, as highlighted by the gold dashed rectangle in panel **a**. (Top) argon gas environment. (Bottom) helium gas environment. The maroon and green dashed framed scalograms correspond to the unstable and stable keyhole cases, respective. The laser spot size was ~77 $\mu$m.

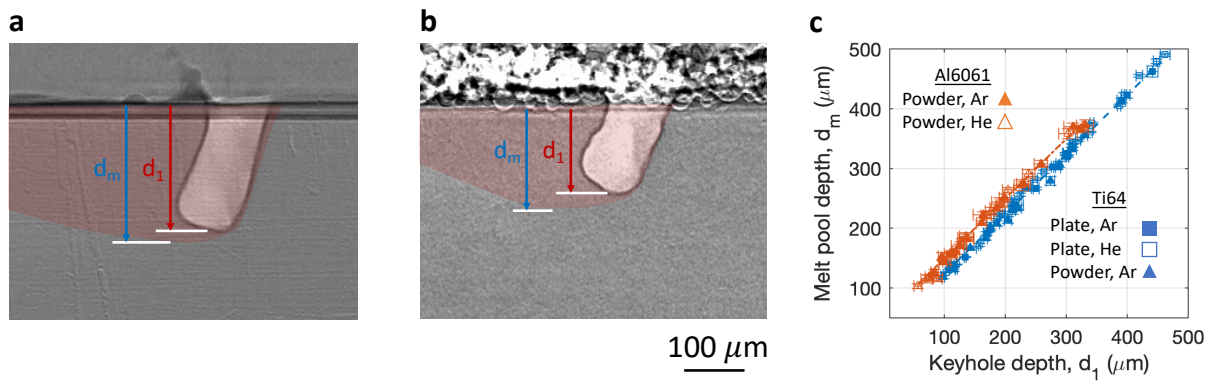

**Fig. 3. Relationship between melt pool depth $d_m$ and keyhole depth $d_1$ in various environment gases, materials, and sample types. a.** Representative x-ray image demonstrating the quantification of melt pool depth ($d_m$) and keyhole depth ($d_1$) in a plate sample. **b.** Representative x-ray illustrating the quantification of melt pool depth ($d_m$) and keyhole depth ($d_1$) in a powder sample. The scale bar applies to both a and b panels, where the melt pool is highlighted in red. **c.** Melt pool depth $d_m$ plotted as a function of keyhole depth $d_1$ across different parameters: environment gas (argon vs helium), materials (Ti-6Al-4V vs Al6061), and sample types (plate vs powder).

*3.3 Melt pool and keyhole morphologies*

In the laser melting process, melt pool and keyhole morphologies are intricately linked. A deeper keyhole leads to more multiple reflections, which tends to "trap" more laser trays within the keyhole, thereby absorbing more energy [28, 29]. Consequently, with more energy conducted to the nearby metal, a larger melt pool is formed. The relationship between keyhole depth ($d_1$) and melt pool depth ($d_m$) is linear, with an identical slope across different gas environments, sample types, or materials. The only variation is the intercept, which is dependent on the materials system (Fig. 3c). Furthermore, keyhole depth exhibits a linear relationship with the tangent of the front keyhole wall angle ($\tan\theta$), regardless of the gas environment (Fig. 4b). This indicates that the environment gas does not affect the underpinning physics of laser melting, as both relationships can be well described using Fabrro's model [30].

However, with the identical P-V combination, the melt pool size in helium is smaller than that in argon (Fig. 5a, Fig. 6, and Supplementary Movies 1-6). This phenomenon explains why helium increases the intrinsic oscillation frequency: a smaller melt pool size means a reduced volume/mass of molten metal, which in turn leads to faster oscillation of the melt pool. The reason for a smaller melt pool in helium is attributed to reduced effective energy absorption, caused by a higher energy loss associated with increased plume convection in helium compared to argon (Appdendix A). This is further supported by the observation of a slower drill rate in helium as opposed to argon (Fig. 4c). Consequently, this leads to a reduction in keyhole depth (Fig. 5b) and front wall angle (Fig. 5c), resulting in less laser ray entrapment, and hence, lower energy absorption. Alternatively, Fig. 5b-c can be interpreted as follow: to achieve the same keyhole depth or front wall angle, more energy must be introduced into the system in a helium environment than in argon. This implies greater energy loss for helium under the same energy density input.

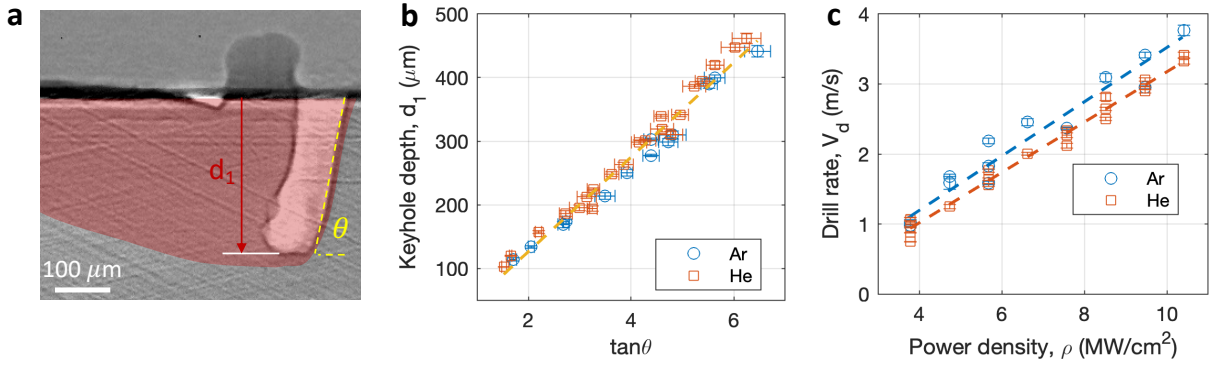
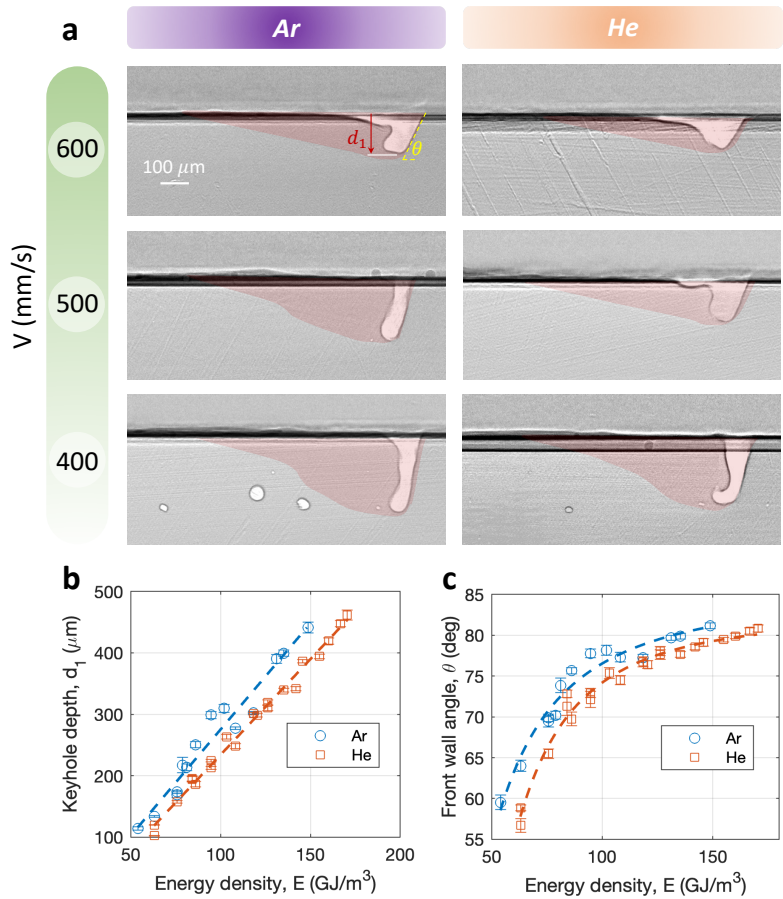

**Fig. 5. Effect of environment gases on keyhole and melt pool morphologies in laser melting of Ti-6Al-4V plates. a.** Tableau of representative x-ray images at a constant laser power (P) of 200 W and varying scan speed (V) for both helium and argon gas environments. The keyhole depth ($d_1$) and front keyhole wall angle ($\theta$) are labelled. The scale bar shown in the first x-ray image is applicable to all x-ray images. In each x-ray image, the melt pool is highlighted in red. The laser spot size (D) was ~77 μm. **b.** Keyhole depth ($d_1$) plotted as a function of energy density ($E = 4P/V\pi D^2$) for both helium and argon gas environments. **c.** Front keyhole wall angle ($\theta$) plotted as a function of energy density (E) for both helium and argon gas environments.
10

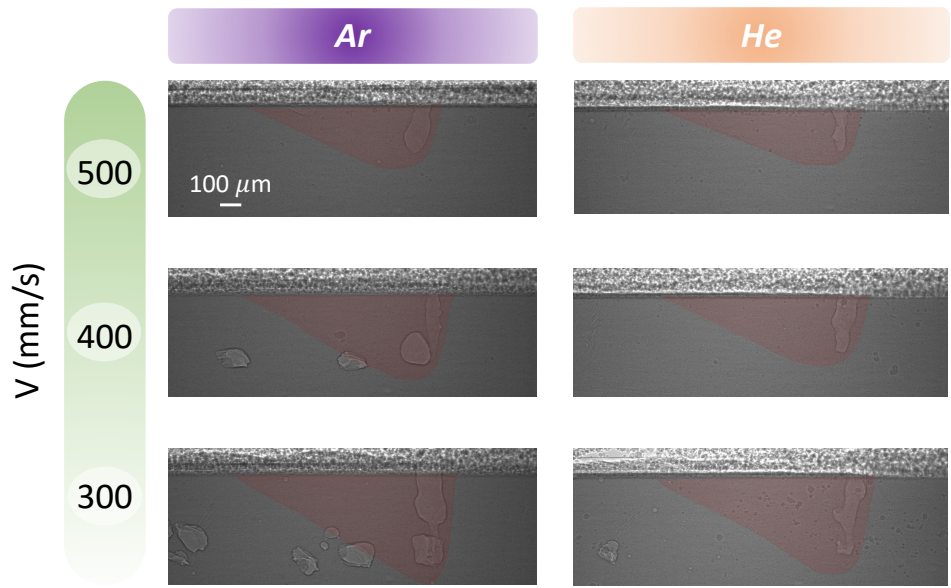

**Fig. 6. Additional example of the effect of environment gases on keyhole and melt pool morphologies, as well as keyhole porosity generation.** In each image, the melt pool is highlighted in red. The sample was Al6061 powder sample with a powder layer of ~ 100 μm. The laser spot size was ~77 μm, and the laser power was 500 W. The scale bar in the first image applies to all images.

*3.4 Surface thermal information*

To calculate the temperature from measured thermal radiation, knowledge of emissivity, which varies with materials, surface qualities, and even temperature itself, is essential but challenging. Nonetheless, in this study, under a well-controlled direct comparison between two gas environments, NIR intensity serve as a viable indicator for comparing cooling curves. This approach is supported by two key factors: firstly, the solidification point is primarily caused by the latent heat of fusion, which is a thermophysical property of the material (Ti-6Al-4V). This is evidenced by the constant intensities of the inflection points of the cooling curves across a various of energy densities (Fig. 7a). Secondly, the effect of the gas type is negligible on the solidification point, suggesting that they do not influence the emissivity (Fig. 7b). Consequently, NIR intensity can be directly used as an indicator to compare the cooling effect between the helium and argon gas environment while controlling the other parameters such as materials, P-V combination, and laser spot size.

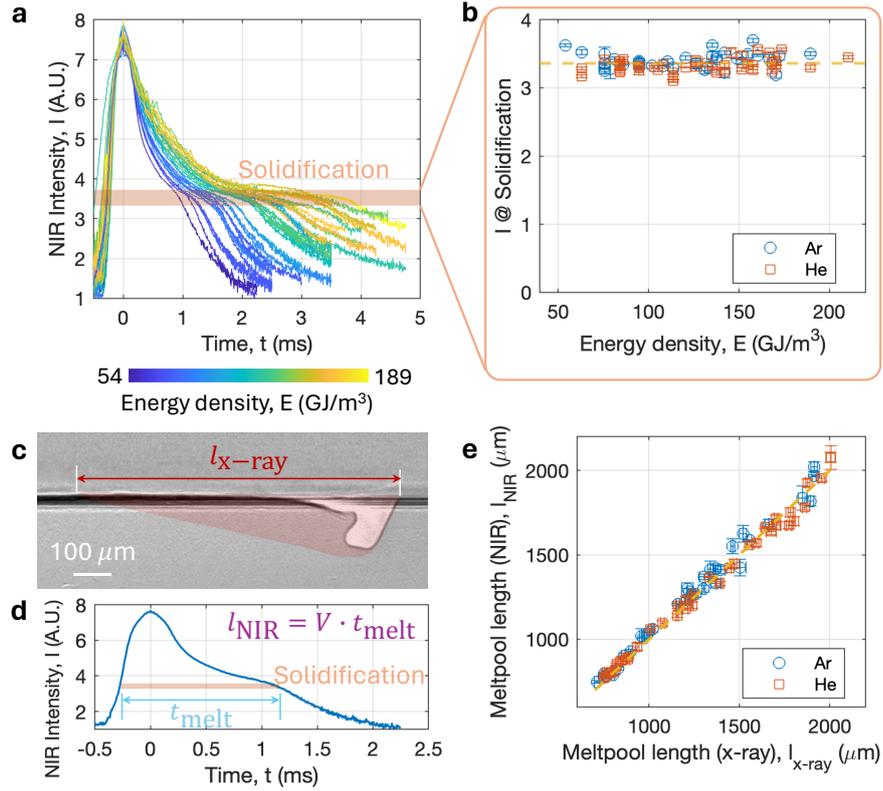

**Fig. 7. Examination of material solidification during laser melting of Ti-6Al-4V plates using NIR intensity. a.** NIR intensity ($I$) as a function of time ($t$) across different energy densities. The solidification point was identified as the turning point on the descending segment of the curve. **b.** Intensity vs energy density plot demonstrating that the intensity at solidification does not vary significantly with either gas environment or energy density. **c.** A representative x-ray image illustrating the quantification of melt pool length. The melt pool was highlighted in red color. **d.** A representative time-series NIR intensity curve demonstrating the identification of melt time ($t_{melt}$) and the estimates of melt pool length from scan speed ($V$) and $t_{melt}$. **e.** Melt pool length quantified from NIR images as a function of melt pool length quantified from x-ray images for both helium and argon gas environments.

The usage of the NIR intensity as a gauge for surface thermal information was validated through an alternative approach. In this method, the high-speed x-ray imaging of the melting process was utilized to provide the melt pool morphology (highlighted in red in Fig. 7c), from which the melt pool length was measured. Subsequently, by identifying the solidification point from the NIR intensity profile, the time period corresponding to how long it takes for the melt pool to pass by a certain point was determined (Fig. 7d). The melt pool length was then calculated from this time period by multiplying it by the laser scan speed ($V$), which is also the melt pool advancing speed. The good agreement of the melt pool length between the x-ray measurement and the NIR measurement over a large range and spanning different gas environments suggests that the use of NIR intensity to evaluate surface thermal information is valid (Fig. 7e).

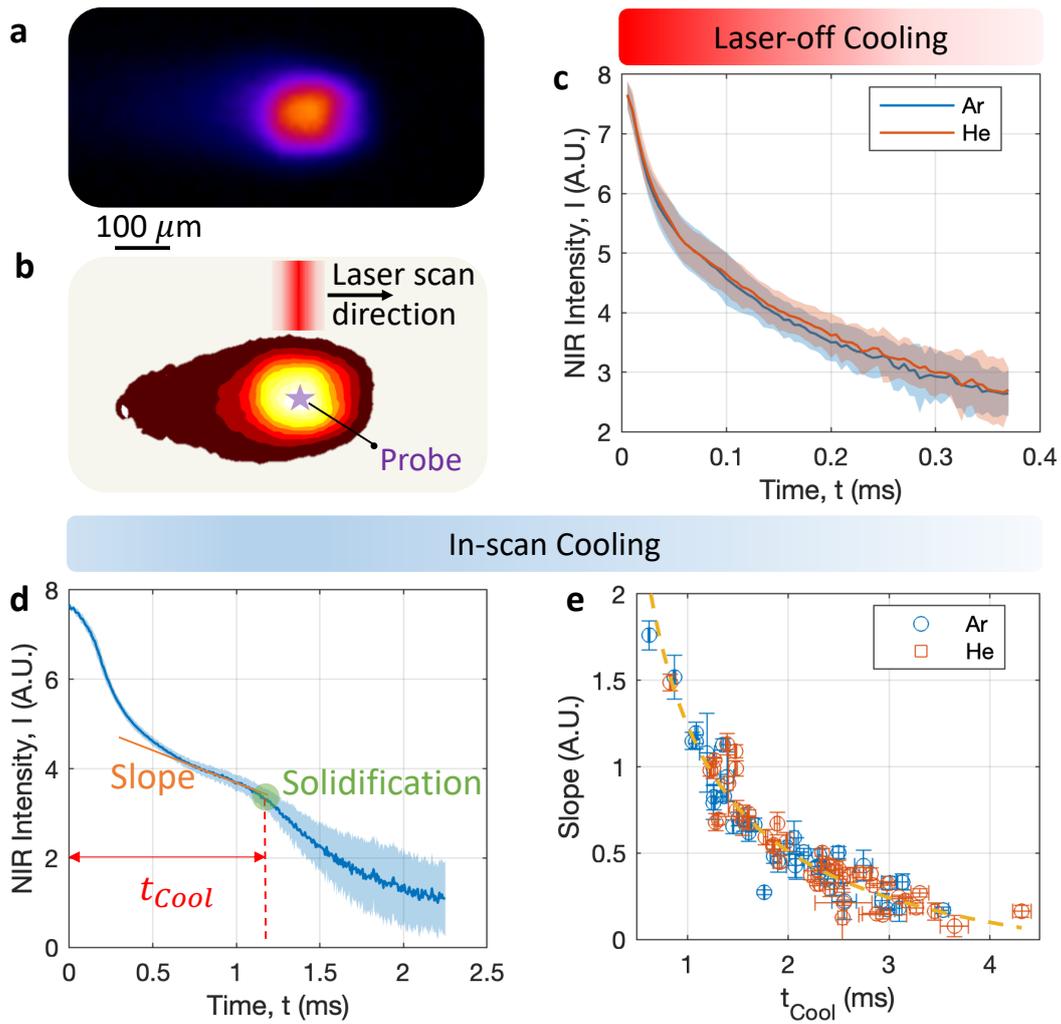

**Fig. 8. Effect of environment gases on cooling rate in laser melting of Ti-6Al-4V plates. a.** A representative raw NIR image of the sample surface. b. Processed NIR image (the same data as panel **a**) with schematic illustrating the extraction of time-series cooling curve from NIR images. The laser (depicted as a red column atop melt pool) moves from left to right. A stationary virtual probe (purple star), used to extract the time-series NIR intensity at a certain location, was marked. The scale bar applies to both panels **a** and **b**. **c.** Comparison of cooling curves between helium and argon gas environments after the laser was deactivated at $t = 0$ (laser-off cooling scenario). In the curve of either gas condition, the error band represents the standard deviations among ~45 P-V combinations. **d.** A representative cooling curve during the laser scan (in-scan cooling scenario). Solidification point, slope at the solidification representing the pseudo cooling rate, and the time from the peak intensity to the solidification ($t_{Cool}$) were highlighted. **e.** Slope as a function of $t_{Cool}$ for helium and argon gas environments. Each point represents a specific P-V combination with error bar illustrating the standard deviation.

*3.5 Cooling curve and cooling rate*

We employed NIR imaging to provide semi-quantitative thermal information of melt pool surface in both high temporal and spatial resolution (Fig. 8a, Supplementary Movies 7-8). Despite the challenges in measuring the absolute temperature due to the unknown emissivity, which varies with materials, surface qualities and even temperature itself, NIR intensity serves as a viable indicator for comparing cooling curves under well-controlled direct comparisons between two gas environments (Fig. 7). A stationary virtual probe was employed to analyze a series of NIR images, extracting the time-series NIR intensity at a specific location (Fig. 8b). We explored the cooling processes of helium and argon in two distinct scenarios: laser-off cooling and in-scan cooling. In the laser-off cooling scenario, we directly compared the cooling processes of helium and argon by deactivating the laser post-scanning. This method enabled us to accurately access the cooling effect of enviroment gases on the solid Ti-6Al-4V with minimum interference from other factors. The energy density during the scanning has negligible impact on the cooling curve, aside from introducing some variations (Fig. 9). Upon considering these variations for statistical significance, it was observed that the helium and argon cool down the sample at nearly identical rates after the laser was turned off (Fig. 8c).

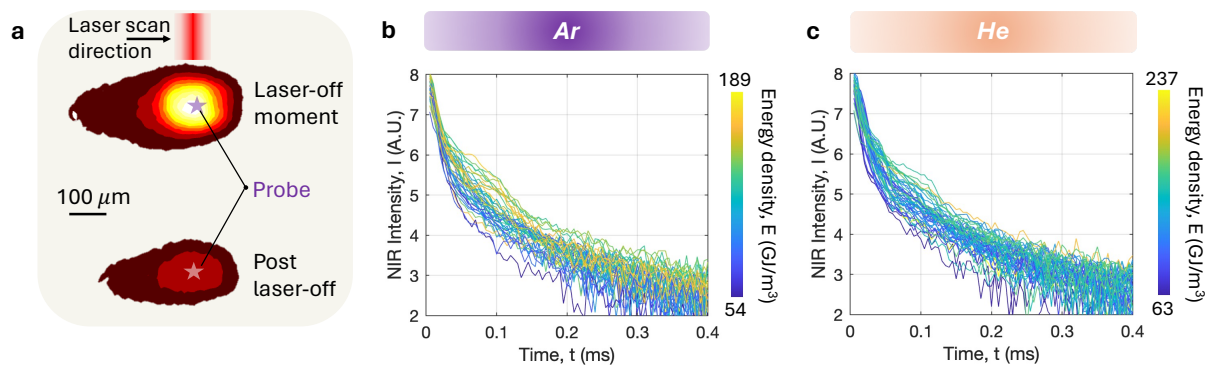

**Fig. 9. Cooling process in laser melting following laser deactivation (laser-off cooling) of Ti-6Al-4V plates. a.** Schematic illustrating the extraction of time-series cooling curve from NIR images in laser-off scenario. The laser (depicted as a red column atop the top melt pool) moves from left to right. The stationary probe (purple star), used to extract the time-series NIR intensity, was positioned along the midline of the melt pool at the location, where the highest intensity occurs at the moment the laser was off ($t = 0$). A representative melt pool at a certain time after the laser-off moment was also shown. **b.** Cooling curves under argon gas conditions across varying energy densities. **c.** Cooling curves under argon gas condition across varying energy densities. The cooling curves appear no obvious patterns regarding the energy densities in neither panel **b** nor **c**. Therefore, the variations in energy density contribute to the quantification of the statistical significance in Fig. 8c.

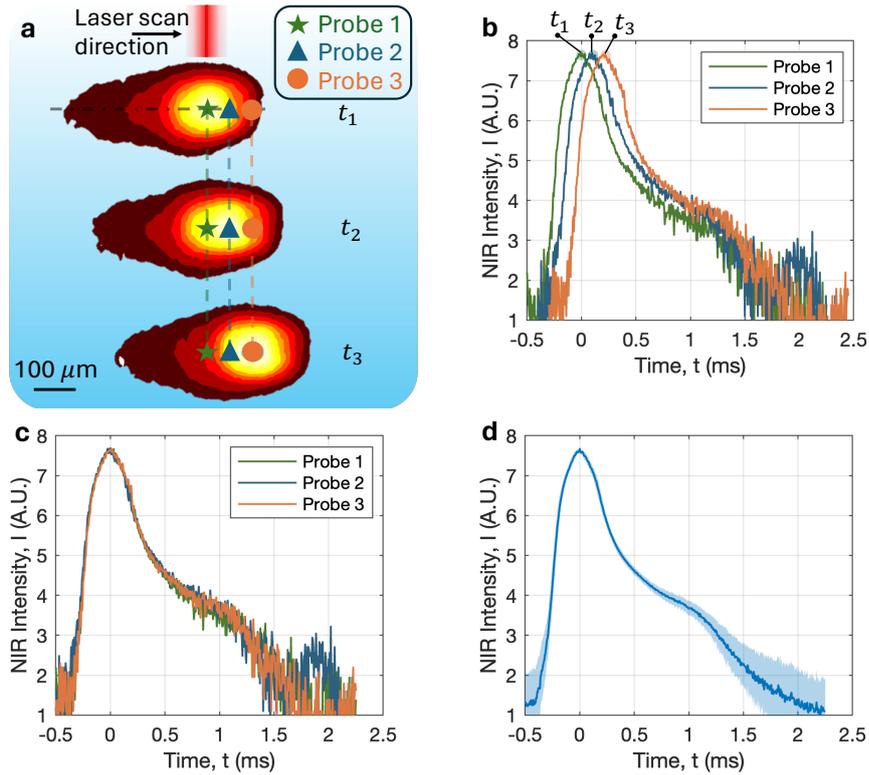

**Fig. 10. NIR intensity profile in laser melting during the scan of Ti-6Al-4V plates. a.** Schematic illustrating the extraction of time-series intensity profile curve from NIR images during the scan. The laser (depicted as a red column atop the top melt pool) moves from left to right. Stationary probes used to extract the time-series NIR intensity, were evenly distributed along the midline of the melt pool (three representative probes shown). The melt pool is shown at different time steps to illustrate the process. **b.** Representative intensity profile extracted from the three probes in panel **a**. The time associated with the peak intensity of each probe corresponds to the respective time steps in panel **a**. **c.** Aligned intensity profiles from panel **b**, adjusted so their peak intensities occurs at $t = 0$. **d.** Representative averaged intensity profile from ~100 probes with the error band representing standard deviation among the probes. The in-scan cooling curve in Fig. 4d represents the segment of the intensity profile following $t = 0$.

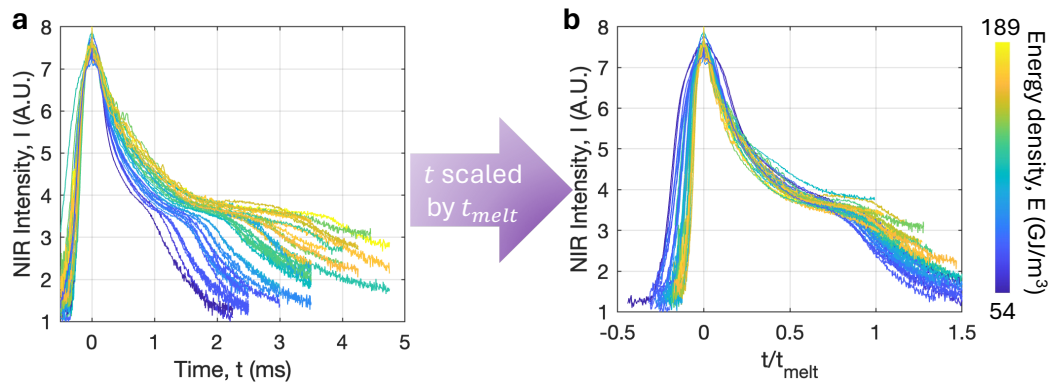

**Fig. 11. Scaling of time-series NIR intensity curves. a.** Time-series NIR intensity curves across varying energy densities before scaling. **b.** Scaled time-series NIR intensity curves across varying energy densities by $t_{melt}$ showing all the curves collapse to a single curve.

In the in-scan cooling scenario, the temperature at the virtual probe does not drop sharply right after the laser passes through as it would in the laser-off cooling scenario; instead, it takes a longer time to cool down because the laser still provides heat even after it surpasses the probe (Fig. 8d). For the same P-V combination, the thermal history of the melt pool at different locations probed along the scan path is almost the same, and hence a representative thermal history can be calculated with statistical meaning across the scanning period (Fig. 10). The NIR intensity at the solidification point, identified as the inflection point on the cooling curve [31], does not vary much with different P-V conditions nor gas environments (Fig. 7a-b). This is because the solidification point is primarily caused by the latent heat of fusion, which is a thermodynamic parameter of the material [32, 33]. The pseudo cooling rate is calculated as the slope of the cooling curve near the solidification point; the time between the peak intensity and the solidification point is defined as cooling period (Fig. 8d). We discovered that the cooling rate decreases with increasing cooling period (Fig. 8e). More importantly, helium does not have a noticeable influence on the cooling rate compared to argon. It is interesting to note that the NIR intensity curve with a higher laser energy density tends to have a longer cooling period. After normalization of the time by the cooling period, all NIR intensity curves collapse (Fig. 11).

*3.6 Effect of environment gases on powder entrainment and melt track morphology*

The high-velocity vapor plume arising from the vapor depression causes a relatively low-pressure region near the rear-top of the keyhole. The difference between the ambient pressure and the low-pressure region drives the otherwise stationary metal powder backward and upward, along the direction of plume ejection. This entrainment of the powder is called powder entrainment. The extent of powder entrainment is directly related to the vapor plume velocity [13]. With a heavier density, the argon gas has a relatively slow vapor plume velocity, and hence weaker powder entrainment (Fig. 12e, Supplementary Movie 9). In comparison, with a much lighter density, the helium gas has a much faster vapor plume velocity and corresponding stronger powder entrainment (Fig. 12e, Supplementary Movie 10).

No obvious effect of different environment gases has been found on the total depth of the melt tracks from the ex situ microscopy of the cut cross-section of the as-built sample [15]. In our in situ x-ray imaging study, however, we have discovered that the negligible effect of gas on the total depth (Fig. 12a, d) was a result of two perceivable and distinct effects. First, helium has a much stronger powder entrainment than argon, introducing more metal powder to the melt region and form a greater melt pool height, measured as the vertical distance from the top of the substrate to the top of the melt track. This is evidenced by a greater melt pool height in helium than that in argon (Fig. 12c) [15]. Second, helium has a much faster vapor velocity and hence a larger energy dissipation leading to a relatively smaller effective energy deposited in the melt pool compared to argon, as explained in the previous section. This is supported by the shallower melt pool depth in the helium gas environment than that in the argon gas environment (Fig. 12b) [15]. In summary, despite the similar total depth of the melt track between argon and helium, the effect of environment gases on the constituent melt pool depth and height is very observable, and mechanism behind it is quite different.

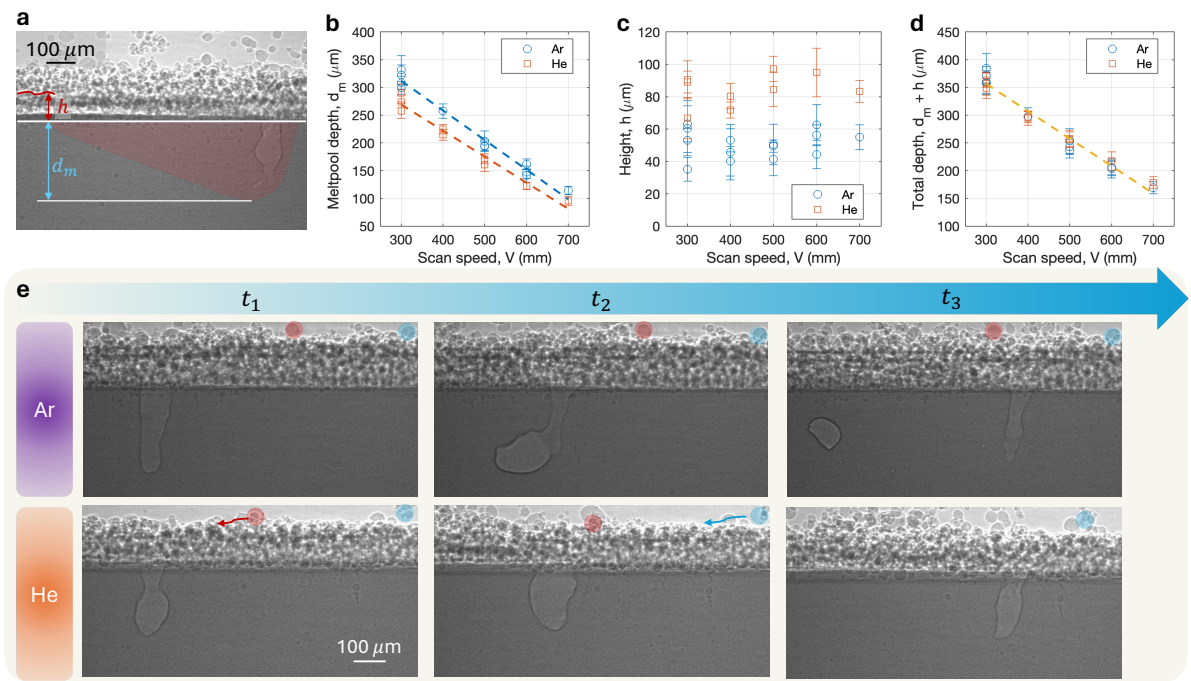

**Fig. 12. Effect of environment gases on powder entrainment and melt track morphology. a.** A representative x-ray image demonstrating the quantification of the melt pool depth ($d_m$) and melt track height (h), measured from the top surface of the substrate to the top of the solidified surface. The melt pool was highlighted in red color. The laser moved from left to right. **b.** Melt pool depth as a function of scan speed (V) for both helium and argon gas environment. **c.** Melt track height as a function of scan speed for both helium and argon gas environment. **d.** Total depth ($d_m$ + h) as a function of scan speed for both helium and argon gas environment. The error bars represent standard deviations. **e.** A series of representative x-ray images captured at three different timesteps ($t_1$ to $t_3$) demonstrating the entrainment of powder in helium environment. Two representative powders were tracked and highlighted in red and blue across the images to illustrate their movements. The scale bar in the left bottom image applies to all images. The material was Al6061. The laser spot size was ~77 μm.

## 4. Discussion

In this study, we investigated the effect of environment gases on the LPBF process through operando experiments (Fig. 1). These experiments allowed for simultaneous characterization of the subsurface melting dynamics and surface thermal signatures of the printing processes. The high-temporal and high-spatial resolutions of both x-ray and NIR imaging provide a holistic view and quantitative analysis of the intricate phenomenon in LPBF under helium environment.

Through the x-ray imaging, we discovered that in the P-V space, helium shifts the porosity boundary to the left compared to the argon environment, yielding a larger processing zone of stable keyhole conditions (Fig. 2a). We also found that helium reduced keyhole pore size compared to argon under the same P-V condition. Both findings suggest that helium stabilizes the keyhole, hence mitigating the generation of keyhole porosity. This was further evidenced by the wavelet analysis on the transmission regime (Fig. 2b), where helium changed the supposed unstable keyhole condition (if in argon) characterized by multimodal perturbative oscillation into a stable keyhole condition characterized by a single-modal intrinsic oscillation. The reduction in porosity in helium was also discussed in [16] through an ex situ cross-sectioning approach. However, a deep understanding and detailed discussion of the mechanisms behind the effect was missing, possibly due to the limitation of the ex situ characterization method.

We found that the reason for helium changing the keyhole oscillation is the reduced melt pool size in the helium environment (Fig. 5a, Fig. 6, and Supplementary Movies 1-6). A smaller melt pool size suggests a reduced volume/mass of molten metal and therefore faster oscillation of the melt pool. The reason for a smaller melt pool in the helium environment is attributed to reduced effective energy absorption, caused by higher energy loss associated with increased plume convection in helium compared to argon (Appendix A). The plume in helium travels much faster than that in argon environment due to its lighter molecular mass [13, 14, 34]. In our work, the higher energy loss associated with helium is also evidenced by a slower drill rate in the helium environment (Fig. 4c) and a consequent reduction in keyhole depth (Fig. 5b) and front wall angle (Fig. 5c). All lead to less laser ray entrapment and thereby lower effective energy absorption.

In comparison, Traoré et al. [15] argued for unchanged energy coupling in their printing of powder samples based on their observation of the similar total depth of the melt tracks between argon and helium. Despite the similar total depth of the melt tracks between argon and helium (Fig. 12a, d), in this work, the effects of helium on the melt pool depth and height were well observed and different underlying mechanisms were distinguished. First, the much faster vapor velocity and the consequent larger energy dissipation in helium lead to a lower effective energy deposited in the melt pool and, hence, a shallower melt pool depth compared to the argon environment (Fig. 12b). The mechanism of the reduced energy coupling was detailed in the last paragraph and Traoré et al. [15] themselves also reported a shallower melt pool depth in helium. Second, the much stronger powder entrainment in helium than that in argon introduces more metal powder to the melt region and forms a greater height of the melt pool above substrate surface. We found that helium gas environment has a stronger powder entrainment compared to argon, which is due to the lower density of helium and the resulting faster vapor plume velocity. Representative Al6061 powder bed samples are shown in section *3.6 Effect of environment gases on powder entrainment and melt track morphology*, Fig. 12e, and Supplementary Movie 9 to 10.

In addition to synchrotron x-ray imaging, we also employed NIR imaging to provide quantitative thermal information of the melt pool surface (Fig. 8a, Supplementary Movies 7-8). A stationary virtual probe was employed to analyze a series of NIR images, extracting the time-series NIR intensity at a specific location (Fig. 8b). We explored the cooling processes of helium and argon in two distinct scenarios: laser-off cooling and in-scan cooling. We found that in either scenario, the presence of helium does not influence the cooling rate of the laser scan process compared to argon (Fig. 8c-e). The faster cooling rate in helium environment has been assumed in the metal additive manufacturing community to explain other effects, such as changed microstructures [15, 17, 18]. However, few works have directly measured and quantified the cooling effect of environment gases in LPBF process. Our experiment observations differs from the simulation results reported in [17], where a faster cooling of helium was simulated. On first thought, one might think the reason may be that our measurement is on the surface while their simulation was probed at the bottom of the melt pool. But the surface temperature should be more sensitive to the gas environment than inside the sample due to the direct contact of the gas with the sample. Another measurement using x-ray diffraction reported in [18] also suggested a comparable cooling rate between argon and helium. So far, we settled the debate on the fast-cooling effect induced by helium on the sample surface through a systematic in situ measurement and direct comparison between two gases: helium does not cool down the sample surface faster than argon during the LPBF process as is usually thought of.

## 5. Conclusions

In summary, through operando experiments with simultaneous high-speed synchrotron x-ray imaging and NIR imaging, we elucidated the effect of environment gases (argon versus helium) on the LPBF process. The quantitative analyses of the keyhole and melt pool, as well as the cooling behavior of the sample, allowed us to settle some long-lasting debates over of the helium effect. That is, helium does not cool the melt pool and solidified material faster than argon through conduction because of its higher thermal conductivity. Instead, helium causes larger energy loss in keyhole mode melting by promoting vapor plume convection because of its lower density than argon. Our work showed that helium can promote the keyhole stability and mitigate keyhole porosity generation, yet the fundamental physics underpinning keyhole dynamics remain the same as in argon environment. In addition, we clarified the confusion regarding the effects of enviroment gases on the melt pool depth and track height in powder bed samples by discovering two distinct mechanisms: helium reduced melt pool size by causing more energy loss so the melt pool depth inside the substrate is smaller; and at the same time, helium enhances the powder entrainment effect so the melt track height above the substrate is larger.

For many industrial applications, helium is certainly not the best choice in LPBF production given its high cost. However, helium serves as an ideal reference gas for fundamental studies because its density is almost an order of magnitude lower than those of commonly used inert gases. Our study highlights the substantial impact of physical properties of environment gases on the build process. As the community broadens the selection of environment gas (inert or reactive) in LPBF, we hope our work can provide some guidances for better planning and utilization of gases in the metal AM processes to produce high-quality parts.

**Acknowledgements**

The authors thank Alex Deriy at the Advanced Photon Source for the assistance in beamline experiments. This research used resources of the Advanced Photon Source, a U.S. Department of Energy (DOE) Office of Science user facility operated for the DOE Office of Science by Argonne National Laboratory under Contract No. DE-AC02-06CH11357.

**Appendix A: Simple model of the energy loss in helium gas environment**

The energy balance in the metal laser melting process is expressed in the form of power as below [8]:

$$P_{laser} = P_{absorb} + P_{reflect} + P_{convect} + P_{evaporate} + P_{radiate} \qquad (1)$$

Where $P_{laser}$ is the total laser power deposited during the melting process and is equal to the laser power $P$. $P_{absorb}$ is the effective absorbed power by the metal, $P_{reflect}$ the power loss due to reflection, $P_{convect}$ the power loss due to the convection, $P_{evaporation}$ the power loss due to evaporation, and $P_{radiate}$ the power loss due to radiation.

Re-organize the equation, we have

$$P_{absorb} = P_{laser} - P_{reflect} - P_{convect} - P_{evaporate} - P_{radiate} \qquad (2)$$

In argon and helium gas environments, we have

$$P_{absorb}^{Ar} = P_{laser}^{Ar} - P_{reflect}^{Ar} - P_{convect}^{Ar} - P_{evaporate}^{Ar} - P_{radiate}^{Ar} \qquad (3)$$

$$P_{absorb}^{He} = P_{laser}^{He} - P_{reflect}^{He} - P_{convect}^{He} - P_{evaporate}^{He} - P_{radiate}^{He} \qquad (4)$$

We make the following assumptions to simplify the calculation while comparing the effect of gas environment under the same condition: keyhole morphology is a half sphere, the keyhole diameter is the laser spot size, the keyhole size for argon and helium is the same, and the temperature inside the keyhole is the boiling point (Fig. A1). Then we have:

$P_{reflect}^{Ar} = P_{reflect}^{He}$ and $P_{evaporate}^{Ar} = P_{evaporate}^{He}$

In addition, we have $P_{laser}^{Ar} = P_{laser}^{He} = P$

(4)-(3), we have

$$P_{absorb}^{He} - P_{absorb}^{Ar} = (P_{convect}^{Ar} - P_{convect}^{He}) + (P_{radiate}^{Ar} - P_{radiate}^{He}) \qquad (5)$$

Divide both sides of the equation by the laser power, we have

$$(P_{absorb}^{He} - P_{absorb}^{Ar})/P = (P_{convect}^{Ar} - P_{convect}^{He})/P + (P_{radiate}^{Ar} - P_{radiate}^{He})/P \qquad (6)$$

$$P_{radiate} = \iint_\Gamma \sigma_{SB} \varepsilon (T^4 - T_\infty^4) dS \qquad (7)$$

Where $\sigma_{SB}$ is the Stefan-Bolztman constant and $\varepsilon$ the emissivity, $T$ and $T_\infty$ the melt pool and ambient temperature, respectively.

Substituting the computational constants (Tab. A1), we have $P_{radiate}/P =0.016\%$ and is therefore negligible

(6) becomes

$$(P_{absorb}^{He} - P_{absorb}^{Ar})/P = (P_{convect}^{Ar} - P_{convect}^{He})/P \qquad (8)$$

$$P_{radiate} = \iint_{\Gamma'} \rho_g c_{pg} T \mathbf{u} \cdot \mathbf{n}_{\Gamma'} \, dS \qquad (9)$$

Where $\rho_g$ is the gas density, $c_{pg}$ heat capacity, T temperature, u plume velocity vector, $\mathbf{n}_{\Gamma'}$ normal directional vector.

Substituting the gas properties (Tab. A2), we have

$$(P_{absorb}^{He} - P_{absorb}^{Ar})/P = -0.1272$$

Typical absorption rate of the Ti-6Al-4V plate in argon gas environment is ~0.4 [35]. The absorption rate in helium decreases by ~30% compared to that in argon.

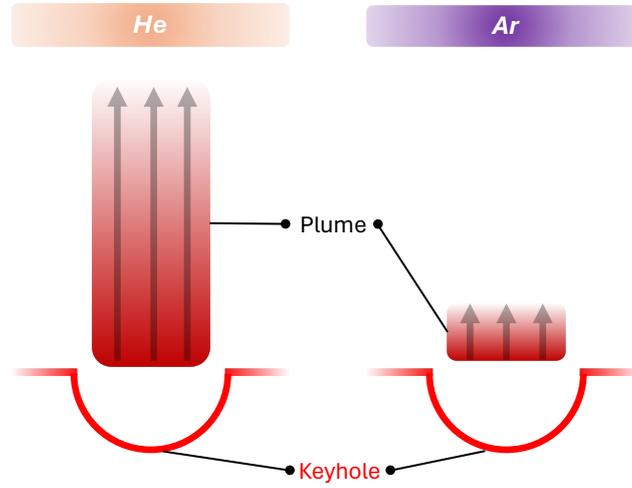

**Fig. A1. Schematic of plume rising from the keyhole in helium and argon gas environment.** The length of the arrow line represents the magnitude of the plume velocity.

**Tab. A1. Computational constants [8].**

| | |
|---|---|
| Stefan-Boltzman constant, $\rho_g$ (kg/m³) | $5.67 \times 10^{-8}$ |
| Emissivity of Ti-6Al-4V | 0.3 |
| Boiling point of Ti-6Al-4V, T (K) | 3133 |
| Ambient temperature, $T_\infty$ (K) | 293 |

**Tab. A2. Properties of Argon and Helium environment gas [8, 13].**

| Properties | Argon | Helium |
|---|---|---|
| Density, $\rho_g$ (kg/m$^3$) | 1.661 | 0.166 |
| Heat capacity, $c_{pg}$ (J/kg·K) | 520.3 | 5193.2 |
| Velocity, u (m/s) | 300 | 1500 |

# References

[1]     T. DebRoy, T. Mukherjee, H. L. Wei, J. W. Elmer, and J. O. Milewski, "Metallurgy, mechanistic models and machine learning in metal printing," *Nature Reviews Materials,* vol. 6, no. 1, pp. 48-68, 2021/01/01 2021, doi: 10.1038/s41578-020-00236-1.

[2]     Y. Chen *et al.*, "In situ X-ray quantification of melt pool behaviour during directed energy deposition additive manufacturing of stainless steel," *Materials Letters,* vol. 286, p. 129205, 2021/03/01/ 2021, doi: https://doi.org/10.1016/j.matlet.2020.129205.

[3]     A. M. Aiden *et al.*, "Dynamics of pore formation during laser powder bed fusion additive manufacturing," *Nature Communications*, vol. 10, no. 1, pp. 1-10 doi: 10.1038/s41467-019-10009-2.

[4]     T. DebRoy *et al.*, "Scientific, technological and economic issues in metal printing and their solutions," *Nature Materials,* vol. 18, no. 10, pp. 1026-1032, 2019/10/01 2019, doi: 10.1038/s41563-019-0408-2.

[5]     T. Sun, W. Tan, L. Chen, and A. Rollett, "In situ/operando synchrotron x-ray studies of metal additive manufacturing," *MRS Bulletin,* vol. 45, no. 11, pp. 927-933, 2020, doi: 10.1557/mrs.2020.275.

[6]     I. Bitharas, N. Parab, C. Zhao, T. Sun, A. D. Rollett, and A. J. Moore, "The interplay between vapour, liquid, and solid phases in laser powder bed fusion," *Nature Communications,* vol. 13, no. 1, p. 2959, 2022/05/26 2022, doi: 10.1038/s41467-022-30667-z.

[7]     S. M. H. Hojjatzadeh *et al.*, "Pore elimination mechanisms during 3D printing of metals," *Nature communications,* vol. 10, no. 1, pp. 1-8, 2019.

[8]     Z. Gan *et al.*, "Universal scaling laws of keyhole stability and porosity in 3D printing of metals," *Nature Communications,* vol. 12, no. 1, p. 2379, 2021/04/22 2021, doi: 10.1038/s41467-021-22704-0.

[9]     M. Qu *et al.*, "Controlling process instability for defect lean metal additive manufacturing," *Nature Communications,* vol. 13, no. 1, p. 1079, 2022/02/28 2022, doi: 10.1038/s41467-022-28649-2.

[10]    Y. Huang *et al.*, "Keyhole fluctuation and pore formation mechanisms during laser powder bed fusion additive manufacturing," *Nature Communications,* vol. 13, no. 1, p. 1170, 2022/03/04 2022, doi: 10.1038/s41467-022-28694-x.

[11]    S. K. H. Everton, Matthias; Stravroulakis, Petros; Leach, Richard K.; Clare, Adam T., "Review of in-situ process monitoring and in-situ metrology for metal additive manufacturing," *Materials and Design,* vol. 95, pp. 431-445, 2016.

[12]     M. Grasso, A. Remani, A. Dickins, B. M. Colosimo, and R. K. Leach, "In-situ measurement and monitoring methods for metal powder bed fusion: an updated review," *Measurement Science and Technology,* vol. 32, no. 11, p. 112001, 2021/07/15 2021, doi: 10.1088/1361-6501/ac0b6b.

[13]     M. A. Stokes, S. A. Khairallah, A. N. Volkov, and A. M. Rubenchik, "Fundamental physics effects of background gas species and pressure on vapor plume structure and spatter entrainment in laser melting," *Additive Manufacturing,* vol. 55, p. 102819, 2022/07/01/ 2022, doi: https://doi.org/10.1016/j.addma.2022.102819.

[14]     Z. Li, G. Yu, X. He, Z. Gan, and W. K. Liu, "Vapor-induced flow and its impact on powder entrainment in laser powder bed fusion," *Materials Today Communications,* vol. 36, p. 106669, 2023/08/01/ 2023, doi: https://doi.org/10.1016/j.mtcomm.2023.106669.

[15]     S. Traore *et al.*, "Influence of gas atmosphere (Ar or He) on the laser powder bed fusion of a Ni-based alloy," *Journal of Materials Processing Technology,* vol. 288, p. 116851, 2021/02/01/ 2021, doi: https://doi.org/10.1016/j.jmatprotec.2020.116851.

[16]     S. Traoré *et al.*, "Influence of gaseous environment on the properties of Inconel 625 L-PBF parts," *Journal of Manufacturing Processes,* vol. 84, pp. 1492-1506, 2022/12/01/ 2022, doi: https://doi.org/10.1016/j.jmapro.2022.11.023.

[17]     H. Amano *et al.*, "Effect of a helium gas atmosphere on the mechanical properties of Ti-6Al-4V alloy built with laser powder bed fusion: A comparative study with argon gas," *Additive Manufacturing,* vol. 48, p. 102444, 2021/12/01/ 2021, doi: https://doi.org/10.1016/j.addma.2021.102444.

[18]     C. Pauzon *et al.*, "Effect of helium as process gas on laser powder bed fusion of Ti-6Al-4V studied with operando diffraction and radiography," *European Journal of Materials,* vol. 2, no. 1, pp. 422-435, 2022/12/31 2022, doi: 10.1080/26889277.2022.2081622.

[19]     A. Caballero, W. Suder, X. Chen, G. Pardal, and S. Williams, "Effect of shielding conditions on bead profile and melting behaviour in laser powder bed fusion additive manufacturing," *Additive Manufacturing,* vol. 34, p. 101342, 2020/08/01/ 2020, doi: https://doi.org/10.1016/j.addma.2020.101342.

[20]     C. Pauzon, P. Forêt, E. Hryha, T. Arunprasad, and L. Nyborg, "Argon-helium mixtures as Laser-Powder Bed Fusion atmospheres: Towards increased build rate of Ti-6Al-4V," *Journal of Materials Processing Technology,* vol. 279, p. 116555, 2020/05/01/ 2020, doi: https://doi.org/10.1016/j.jmatprotec.2019.116555.

[21]     C. Pauzon *et al.*, "Residual stresses and porosity in Ti-6Al-4V produced by laser powder bed fusion as a function of process atmosphere and component design," *Additive*

*Manufacturing,* vol. 47, p. 102340, 2021/11/01/ 2021, doi: https://doi.org/10.1016/j.addma.2021.102340.

[22] C. Zhao *et al.*, "Real-time monitoring of laser powder bed fusion process using high-speed X-ray imaging and diffraction," *Scientific Reports,* vol. 7, 2017.

[23] N. D. Parab *et al.*, "Ultrafast X-ray imaging of laser–metal additive manufacturing processes," *Journal of synchrotron radiation,* vol. 25, no. 5, pp. 1467-1477, 2018.

[24] Z. Ren *et al.*, "Machine learning–aided real-time detection of keyhole pore generation in laser powder bed fusion," *Science,* vol. 379, no. 6627, pp. 89-94, 2023/01/06 2023, doi: 10.1126/science.add4667.

[25] C. Zhao *et al.*, "Critical instability at moving keyhole tip generates porosity in laser melting," *Science,* vol. 370, no. 6520, p. 1080, 2020, doi: 10.1126/science.abd1587.

[26] M. Hamidi Nasab *et al.*, "Harmonizing sound and light: X-ray imaging unveils acoustic signatures of stochastic inter-regime instabilities during laser melting," *Nature Communications,* vol. 14, no. 1, p. 8008, 2023/12/05 2023, doi: 10.1038/s41467-023-43371-3.

[27] D. Guirguis, C. Tucker, and J. Beuth, "Accelerating process development for 3D printing of new metal alloys," *Nature Communications,* vol. 15, no. 1, p. 582, 2024/01/17 2024, doi: 10.1038/s41467-024-44783-5.

[28] C. Zhao *et al.*, "Bulk-Explosion-Induced Metal Spattering During Laser Processing," *Physical Review X,* vol. 9, no. 2, p. 021052, 06/14/ 2019, doi: 10.1103/PhysRevX.9.021052.

[29] S. A. Khairallah, A. T. Anderson, A. Rubenchik, and W. E. King, "Laser powder-bed fusion additive manufacturing: Physics of complex melt flow and formation mechanisms of pores, spatter, and denudation zones," *Acta Materialia,* vol. 108, pp. 36-45, 2016/04/15/ 2016, doi: https://doi.org/10.1016/j.actamat.2016.02.014.

[30] R. Fabbro and K. Chouf, "Keyhole modeling during laser welding," *Journal of Applied Physics,* vol. 87, no. 9, pp. 4075-4083, 2000, doi: 10.1063/1.373033.

[31] B. Lane, S. Moylan, E. P. Whitenton, and L. Ma, "Thermographic measurements of the commercial laser powder bed fusion process at NIST," *Rapid Prototyping Journal,* vol. 22, no. 5, pp. 778-787, 2016, doi: 10.1108/RPJ-11-2015-0161.

[32] P. A. Hooper, "Melt pool temperature and cooling rates in laser powder bed fusion," *Additive Manufacturing,* vol. 22, pp. 548-559, 2018.

[33] W. Kurz, C. Bezençon, and M. Gäumann, "Columnar to equiaxed transition in solidification processing," *Science and Technology of Advanced Materials,* vol. 2, no. 1, pp. 185-191, 2001/01/01 2001, doi: 10.1016/S1468-6996(01)00047-X.

[34]	P. Bidare, I. Bitharas, R. M. Ward, M. M. Attallah, and A. J. Moore, "Laser powder bed fusion in high-pressure atmospheres," *The International Journal of Advanced Manufacturing Technology,* vol. 99, no. 1, pp. 543-555, 2018/10/01 2018, doi: 10.1007/s00170-018-2495-7.

[35]	A. Khairallah Saad *et al.*, "Controlling interdependent meso-nanosecond dynamics and defect generation in metal 3D printing," *Science,* vol. 368, no. 6491, pp. 660-665, 2020/05/08 2020, doi: 10.1126/science.aay7830.